\documentclass[twocolumn,american,english,aps,manuscript,prl,10pt]{revtex4-1}
\usepackage[T1]{fontenc}
\usepackage[latin9]{inputenc}
\usepackage{graphicx}

\usepackage{hyperref}
\hypersetup{
 hidelinks}
\usepackage{color}
\makeatletter

\@ifundefined{textcolor}{}
{%
 \definecolor{BLACK}{gray}{0}
 \definecolor{WHITE}{gray}{1}
 \definecolor{RED}{rgb}{1,0,0}
 \definecolor{GREEN}{rgb}{0,1,0}
 \definecolor{BLUE}{rgb}{0,0,1}
 \definecolor{CYAN}{cmyk}{1,0,0,0}
 \definecolor{MAGENTA}{cmyk}{0,1,0,0}
 \definecolor{YELLOW}{cmyk}{0,0,1,0}
}

\makeatother

\usepackage{babel}
\begin{document}

\preprint{MIT-CTP/4700}

\title{Rapidity dependence in holographic heavy ion collisions}

\author{Wilke van der Schee$^1$ and Bj\"orn Schenke$^2$}

\affiliation{$^1$ Center for Theoretical Physics, MIT, Cambridge, MA 02139, USA}
\affiliation{$^2$ Physics Department, Bldg. 510A,
Brookhaven National Laboratory, Upton, NY 11973, USA}

\selectlanguage{english}%
\begin{abstract}
We present an attempt to closely mimic the initial stage of heavy ion collisions
within holography, assuming a decoupling of longitudinal and transverse
dynamics in the very early stage. We subsequently evolve the obtained
initial state using state-of-the-art hydrodynamic simulations,
and compare results to experimental data. We present
results for charged hadron pseudo-rapidity spectra and directed and elliptic flow as
functions of pseudo-rapidity for $\sqrt{s_{NN}}=200\,{\rm GeV}$ Au-Au and $2.76\,{\rm TeV}$
Pb-Pb collisions. The directed flow interestingly turns out to be quite sensitive
to the viscosity. The results can explain qualitative features
of the collisions, but the
rapidity spectra in our current model is narrower than the experimental data.
\end{abstract}

\maketitle
\textbf{1. Introduction. } Collisions of relativistic heavy ions
have been understood to quickly form a quark-gluon plasma (QGP), which
thereafter evolves according to relativistic hydrodynamics with small
viscosity \cite{Gale:2013da,Heinz:2013th,deSouza:2015ena}. To make this paradigm
more precise it is of crucial importance to understand the early time far-from-equilibrium
stage and be able to accurately compute the initial state of the hydrodynamic
plasma. This is particularly challenging because of the generally non-perturbative nature of
Quantum Chromo Dynamics (QCD). In this Letter we model this initial stage at strong coupling using
holography. We then provide the resulting energy density and flow velocity distributions as
input for the subsequent viscous hydrodynamic evolution.

This first study makes several strong assumptions. In particular, we
describe the entire early stage of the collision in the strong coupling limit. Further,
we work in the canonical holographic theory, $\mathcal{N}=4$
super-Yang-Mills (SYM) theory with a large number of colors. This conformal theory
has no confinement or asymptotic freedom and hence is very different from QCD.
Nevertheless, at energy scales relevant for the early stage of heavy ion collisions
the theories are more similar, which is where we use the SYM theory as an approximation for QCD. Lastly,
this study neglects  any chemical potentials and event-by-event fluctuations.

Previous studies of QGP thermalization at strong coupling notably
include homogeneous \cite{Heller:2012km} and boost-invariant \cite{Chesler:2009cy,Heller:2011ju}
settings, less trivially the collisions of planar shock waves \cite{Chesler:2010bi,Casalderrey-Solana:2013aba,Casalderrey-Solana:2013sxa}
and transverse expansion \cite{vanderSchee:2012qj,vanderSchee:2013pia,Habich:2014jna,Chesler:2015wra}.
It is the purpose of this Letter to combine lessons from these works
to construct the initial conditions for hydrodynamics, in particular
using the fast thermalization \cite{Chesler:2009cy,Heller:2011ju}, a universal rapidity profile \cite{Casalderrey-Solana:2013aba,Casalderrey-Solana:2013sxa} and a
simple formula for transverse flow \cite{vanderSchee:2012qj,Habich:2014jna}. We thereafter employ the \textsc{Music}
viscous relativistic hydrodynamic simulation \cite{Schenke:2010nt,Schenke:2010rr,Schenke:2011bn}
to evolve this profile till freeze-out, after
which we obtain the particle spectra using the Cooper-Frye formalism
\cite{Cooper:1974qi}.

At strong coupling there exists a specific rapidity profile which
is notably different from other approaches \cite{Casalderrey-Solana:2013sxa,vanderSchee:2014qwa},
and we therefore focus on the rapidity dependence of observables,
in particular directed flow, which is non-trivial to reproduce
with initial conditions that model the longitudinal structure \cite{Bozek:2010bi}. Without including full
event-by-event fluctuating initial conditions, it is well-known that
quantitative agreement with experimental data cannot be achieved. 
Nevertheless, comparing our results to experimental data for the pseudo-rapidity 
distributions of charged hadrons, we find that while the profiles are narrower than the experimental data,
agreement is better than expected given the very narrow initial rapidity profile
characteristic of holography \cite{Casalderrey-Solana:2013aba}. Also, we find good quantitative agreement of the directed flow as a function of pseudo-rapidity, both at
RHIC and LHC energies around mid-rapidity. A more complete analysis will therefore be of great interest, as will
be a direct quantitative comparison with models inspired by perturbative
QCD \cite{Gale:2012rq}.

Lastly, we found two important results, likely independent of the holographic 
framework used here. 
Firstly, we found that almost half of the produced entropy
is due to viscous entropy production, even though the viscosity is
small. 
Secondly, the rapidity profile of the directed flow is very
sensitive to the viscosity, and may as such be useful to improve future
estimates of the viscosity of the QGP.

\noindent \textbf{2. The initial state from holography.} Holography
provides a precise mapping between certain strongly coupled quantum
field theories and gravitational theories with one extra dimension.
Here we will use the original and simplest example, where strongly coupled $SU(N_{c})$
$\mathcal{N}=4$ super Yang-Mills theory in 3+1 dimensions is mapped
to a gravitational theory in 4+1 dimensional anti-de-Sitter spacetime.
As has often been done to describe heavy ion collisions,
 the individual ions are \foreignlanguage{american}{modeled}
by gravitational shock waves, which in the Yang-Mills theory correspond
to lumps of energy moving unperturbed at the speed of light.

An important assumption we will be making is the decoupling of the
longitudinal dynamics and the transverse dynamics in the stage before
we use hydrodynamics, which we here take to be the first $0.1\,(0.2)$ fm/c
of the collision at LHC (RHIC). This assumption should hold to high
accuracy as long as the typical transverse structures are larger than
this time. For an average energy density this would even be the case
in systems as small as the ones in p-A collisions, but note that it prevents
us from considering event-by-event fluctuations smaller than
this length. This assumption, however, allows us to split up the holographic
calculation into a longitudinal one with translational symmetry in the transverse
plane, and a transverse calculation with boost invariance.

\begin{figure}
\begin{centering}
\includegraphics[width=9cm]{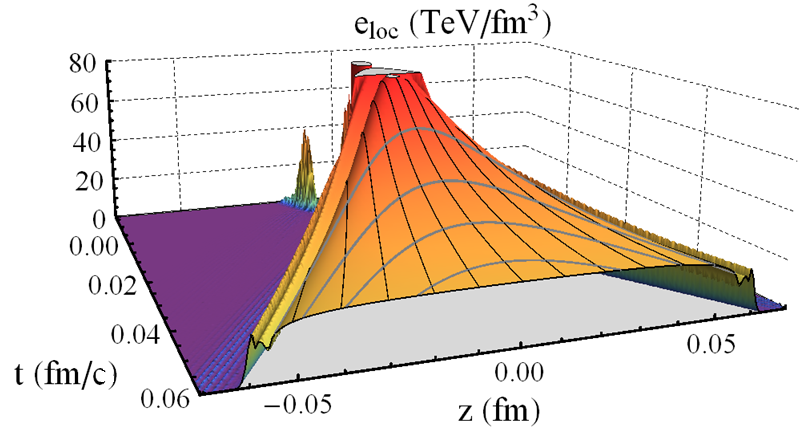}
\par\end{centering}

\protect\caption{We show the local energy density as a function of time and the longitudinal
direction, in the center of mass frame. Units are such that the energy
per transverse area of the initial shock matches the center of a LHC
Pb-Pb collision. Note that this frame will depend on the position
in the transverse plane. The black and grey curves denote stream lines
and constant proper time curves. \label{fig:eloc}}
\end{figure}

\begin{figure*}
\begin{centering}
\includegraphics[width=9cm]{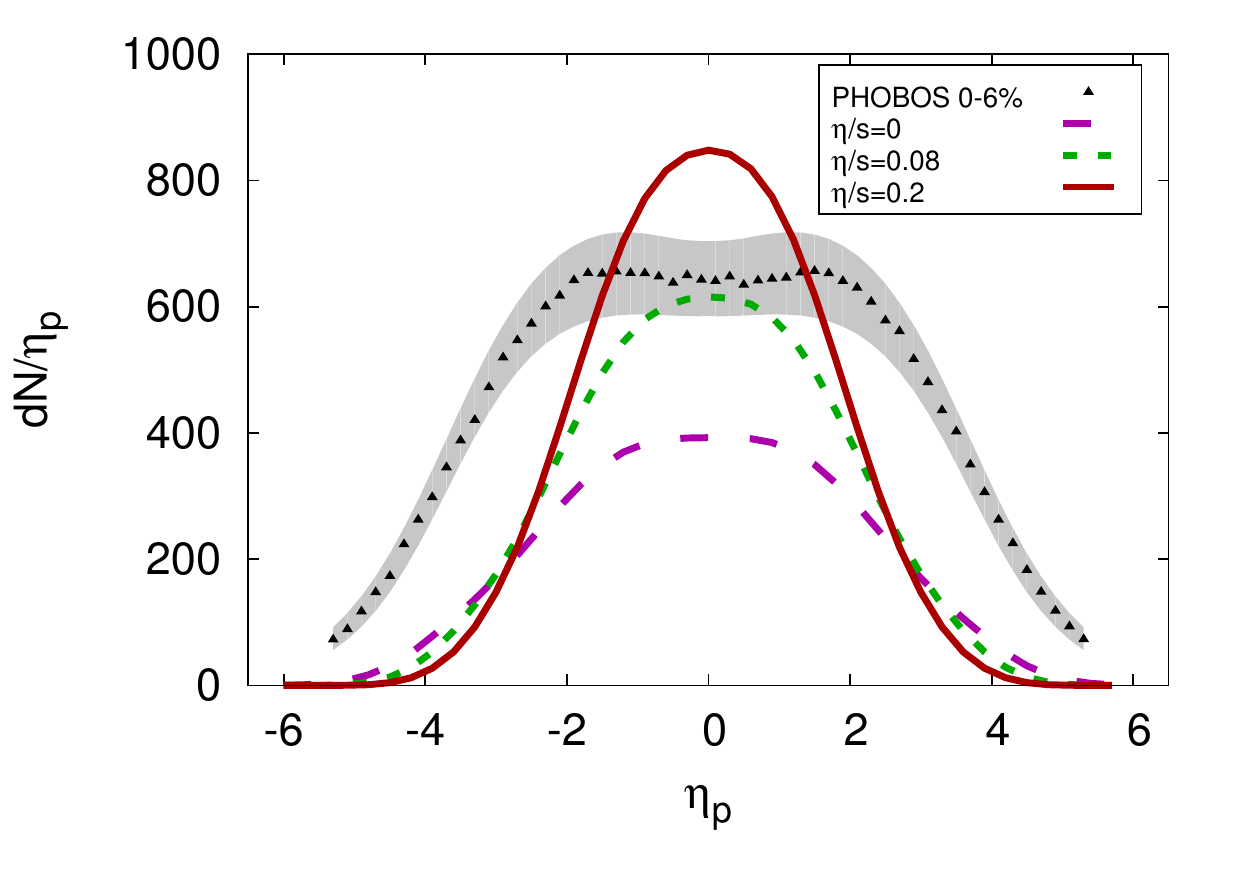}\includegraphics[width=9cm]{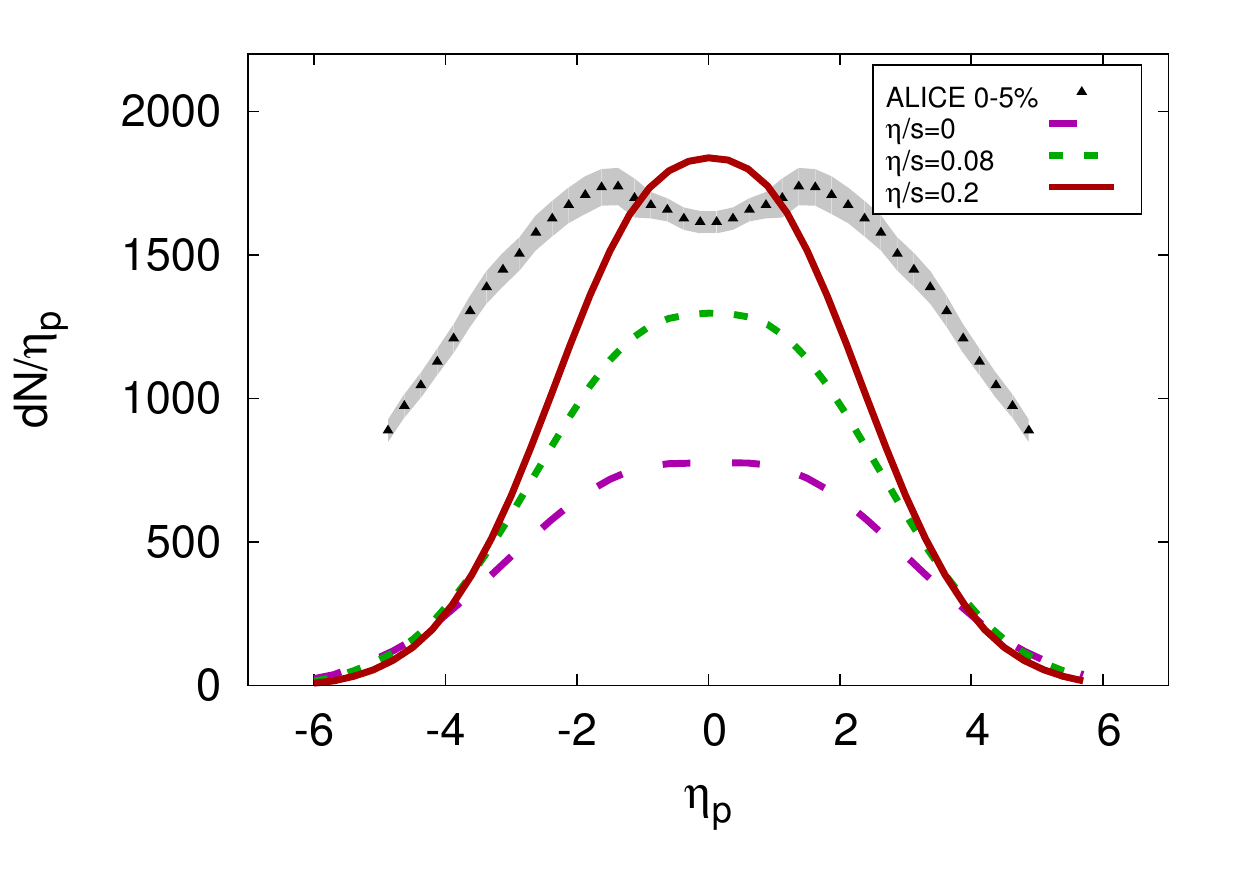}
\par\end{centering}

\protect\caption{Charged hadron multiplicity for top RHIC ($200\,{\rm GeV}$) (left) and LHC ($2.76\,{\rm TeV}$) (right) energies 
as a function of pseudo-rapidity $\eta_p$ for a collision with impact
parameter $1\,{\rm fm}$ with $\eta/s=0$, $0.08$
and $0.2$. In all cases the spectrum is narrower than the experimental
data from the PHOBOS \cite{Alver:2010ck} and ALICE \cite{Abbas:2013bpa} collaborations, respectively.
However, the spectrum is much wider
than the initial state rapidity profile obtained from just holography
(see Fig.\, \ref{fig:eloc} and \cite{Casalderrey-Solana:2013aba}).\label{fig:dndeta}}
\end{figure*}

The longitudinal dynamics has been studied in \cite{Janik:2005zt,Chesler:2010bi,Casalderrey-Solana:2013sxa,vanderSchee:2014qwa, Chesler:2015bba,Chesler:2015fpa}
which led to two main lessons: the plasma thermalizes very fast in
the sense that \emph{viscous} hydrodynamics becomes applicable in
times perhaps as short as 0.05 fm/c at LHC energies. Furthermore,
at LHC energies the temperature is universally approximately constant in the $z$ (beam) direction
at constant time \footnote{Recently this has been computed more precisely in \cite{Chesler:2015fpa}, where it turned out that this observation is slightly modified close to the light cone, causing some small changes at high rapidities. At the level of this initial study this difference is however not significant. This study also computed the normalized $dN/dy$ spectrum for central collisions, which qualitatively agrees with Fig. \ref{fig:dndeta}.}
 (Fig. \ref{fig:eloc}), where importantly the
time is determined in the local center of mass frame (LCOM) of the
local transverse energy densities \cite{Casalderrey-Solana:2013sxa}.
This means it does \emph{not }refer to the nucleon-nucleon center of mass.
This observation is also valid for asymmetric longitudinal profiles \cite{Casalderrey-Solana:2013sxa}, which is relevant for off-central collisions.
Remarkably, even though this temperature profile is not boost invariant
at all, the longitudinal velocity profile shows approximate Bjorken behavior:
$v_{z}=z/t$. 

For the transverse energy densities $\epsilon_L$ and $\epsilon_R$ of the left- and right-moving nuclei we will take an integrated Wood-Saxon distribution \cite{Alver:2008aq}.
In matching the holographic computation to hydrodynamics we then use the following formula
for the energy density \cite{Heller:2007qt}: 
\begin{equation}
\mathcal{E}(t)=\frac{N_{c}^{2}\Lambda^{4}}{2\pi^{2}}\left[\frac{1}{(\Lambda t)^{4/3}}-\frac{2\eta_{0}}{(\Lambda t)^{2}}\right],\label{eq:janik-heller}
\end{equation}
where $\eta_{0}=\frac{1}{\sqrt{2}\,3^{3/4}}$, we take $N_c=1.8$ such that the EOS matches lattice data ($e/T^4 \approx 12$) \cite{Gubser:2008pc,Cheng:2007jq} and $\Lambda$ has to
be extracted from Fig. \ref{fig:eloc} numerically as $\Lambda=0.37\,\epsilon^{1/3}$
\cite{Gubser:2014qua}, with $\epsilon=\sqrt{\epsilon_{L}\epsilon_{R}}$ the center of mass energy
density per transverse area of the collision (all quantities depend on
$x_{\perp}$). 
Eq.\,(\ref{eq:janik-heller}) was originally found as a solution to first order hydrodynamics in a boost invariant context, but it turns out that the formula also works well to describe the energy density of our (non-boost invariant) collision at midrapidity \cite{Chesler:2013lia}. There is no $z$ dependence since the local energy density is approximately constant at constant $t$ (Fig.\,\ref{fig:eloc}) and the dependence on the transverse coordinates is entirely contained in the transverse energy densities.
The rapidity
profile then follows by converting proper time and rapidity to normal
time, where as described above we have to boost to the local center
of mass frame. We hence use $t=\tau\,\cosh(y+\delta y$), with $y$ the spacetime rapidity
and $\delta y=\frac{1}{2}\log(\epsilon_{L}/\epsilon_{R})$ the shift
to the LCOM \cite{Casalderrey-Solana:2013sxa}. 

When the longitudinal width of the incoming nuclei is large enough,
the high energy regime plotted in Fig.\,\ref{fig:eloc} is no longer
applicable \cite{Casalderrey-Solana:2013aba,Chesler:2015lsa}. For
RHIC energies this is a significant effect, and it was found that
the rapidity profile is approximately $30\%$ narrower and $83\%$
higher as compared to the high energy regime described above \cite{Casalderrey-Solana:2013aba}. We incorporated this by
a simple rescaling of the hydrodynamic initial data.

By causality the transverse profile of the energy density does not
change much in shape in this very early stage. A non-trivial transverse
flow does develop \cite{Vredevoogd:2008id}, which has been studied holographically in \cite{vanderSchee:2012qj,vanderSchee:2013pia}.
These works found that the transverse fluid velocity is proportional
to the transverse gradient of the energy density, with a numerical
coefficient extracted in \cite{Habich:2014jna}:
\begin{equation}
v_{i}=-0.33\,\tau\,(\partial_{i}e)/e,\label{eq:transverseflow}
\end{equation}
with $i=x\text{ or }y$, $\tau$ the proper time, and $e$ the local energy density, now
depending on all spatial coordinates.

We can now proceed to the construction of the complete initial condition.
Given two colliding objects, having their respective initial energy
profiles $\epsilon_{L}$ and $\epsilon_{R}$ as a function of the
transverse plane, we construct for all transverse coordinates the
relevant rapidity and energy of the LCOM. We then map the holographic
longitudinal profile (using Eq. (\ref{eq:janik-heller})) to obtain the energy density as a function of
rapidity at a fixed initial proper time $\tau_{\rm ini}=0.1\,(0.2)$ fm/c
at LHC (RHIC). Once we have the local energy density we obtain the
local transverse velocity from Eq.\,(\ref{eq:transverseflow}). 

In the approximation described above holography provides
the complete energy density, in principle only requiring the transverse
energy densities. Nevertheless, we found that this approach overestimated
the total multiplicity, possibly because of not including fluctuations
and the assumption of very strong coupling. For this reason we divided
the energy density by a factor of 20 (6) for the top LHC (RHIC) energies so
that charged hadron multiplicities would be close to the experimental data,
which we will address again in the discussion.

Having obtained the energy density and the local fluid velocity we
proceed by using them to initialize the relativistic viscous fluid dynamic simulation 
\textsc{Music} \cite{Schenke:2010nt,Schenke:2010rr,Schenke:2011bn}. 
We set the initial viscous stress tensor to zero and will use various constant
values of the shear viscosity to entropy density
ratio $\eta/s$.

For the equation of state we use the 
the parametrization ``s95p-v1'' from \cite{Huovinen:2009yb}, obtained from interpolating between lattice data and a hadron resonance gas.

At a constant freeze-out
temperature of $150\,{\rm MeV}$ the fluid is converted to particles using
the Cooper-Frye algorithm \cite{Cooper:1974qi}, after which resonance decays are computed,
including all resonances of energy $2\,{\rm GeV}$ or less.

\noindent \textbf{3. Results and discussion. } We present the
resulting charged hadron spectra as a function of pseudo-rapidity
 in Fig.\,\ref{fig:dndeta} for top RHIC ($\sqrt{s}=200\,{\rm GeV}$) and LHC ($\sqrt{s}=2760\,{\rm GeV}$) energy collisions.
We used an impact parameter of 1 fm to simulate central collisions.
The rapidity spectrum for RHIC and LHC collisions are too narrow, but 
the relative increase in width from RHIC to LHC is similar to the data. Also, it is noteworthy that the rapidity spectra found here are much wider than the holographic
initial profile of width 0.9 found in \cite{Casalderrey-Solana:2013aba}, stressing the importance of hydrodynamic evolution.

We note that the previously observed \cite{Schenke:2011bn} effect of viscosity, namely the reduction of the effective longitudinal pressure 
and correspondingly smaller multiplicities at larger rapidity are also visible in Fig.\,\ref{fig:dndeta}.

\begin{figure}
\begin{centering}
\includegraphics[width=9cm]{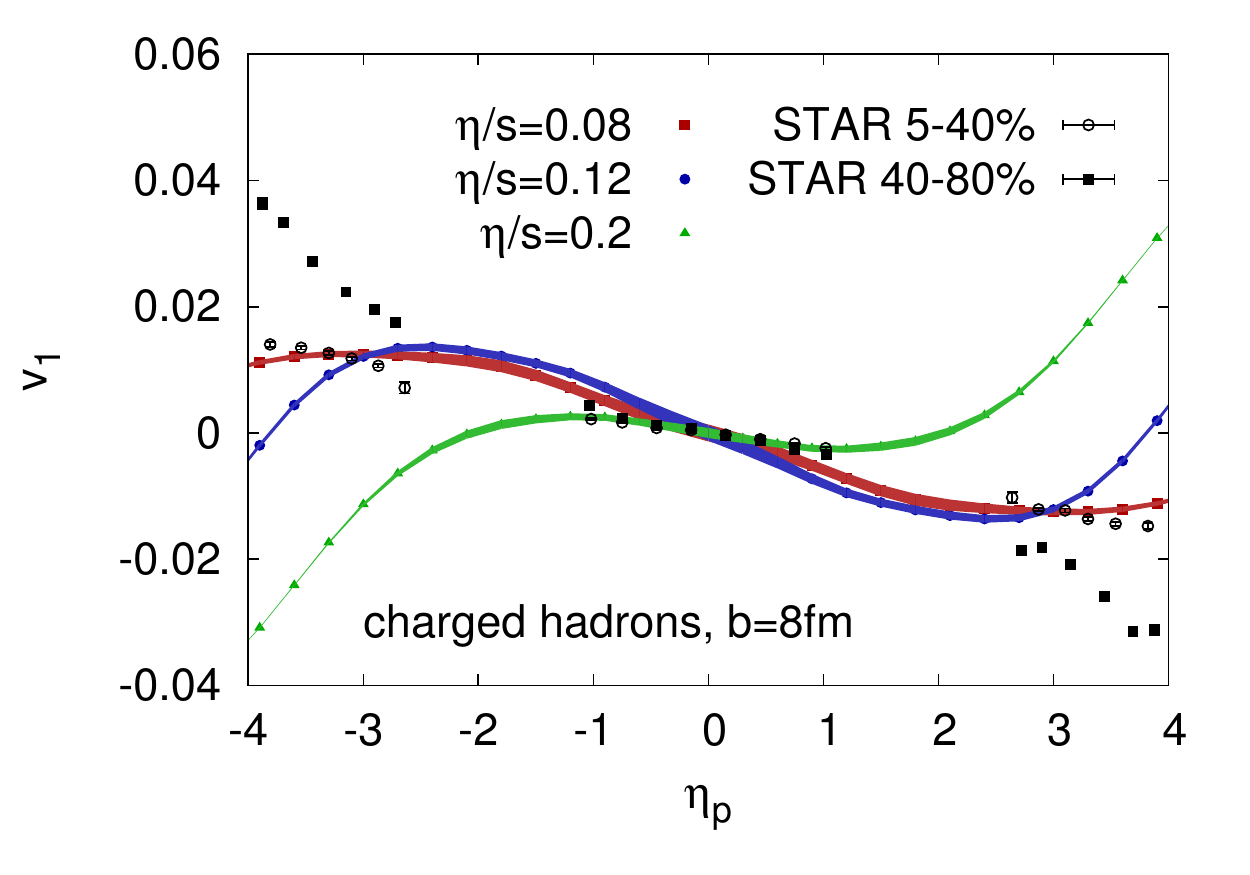}
\par\end{centering}

\protect\caption{Directed flow $v_{1}$ as a function of pseudo-rapidity
$\eta_p$ for 200 GeV collisions at RHIC with $b=8\,{\rm fm}$ (approximately 35\% centrality)
using $\eta/s= 0.08$, $0.12$, and $0.2$.
While a direct comparison with experimental data is difficult
without a proper event-by-event analysis, the shape around midrapidity
matches STAR data for 5-40\% and 40-80\% centrality classes \cite{Abelev:2008jga} quantitatively around  mid-rapidity. The results for large rapidity are uncertain, because there $dN/d\eta_p$ is in disagreement   (Fig. \ref{fig:dndeta}).
This observable is quite sensitive
to the viscosity, which is interesting in its own right. The bands describe uncertainties in $v_1$ stemming from numerical inaccuracies in the freeze-out surface finding.\label{fig:v1}}
\end{figure}

\begin{figure}
\begin{centering}
\includegraphics[width=9cm]{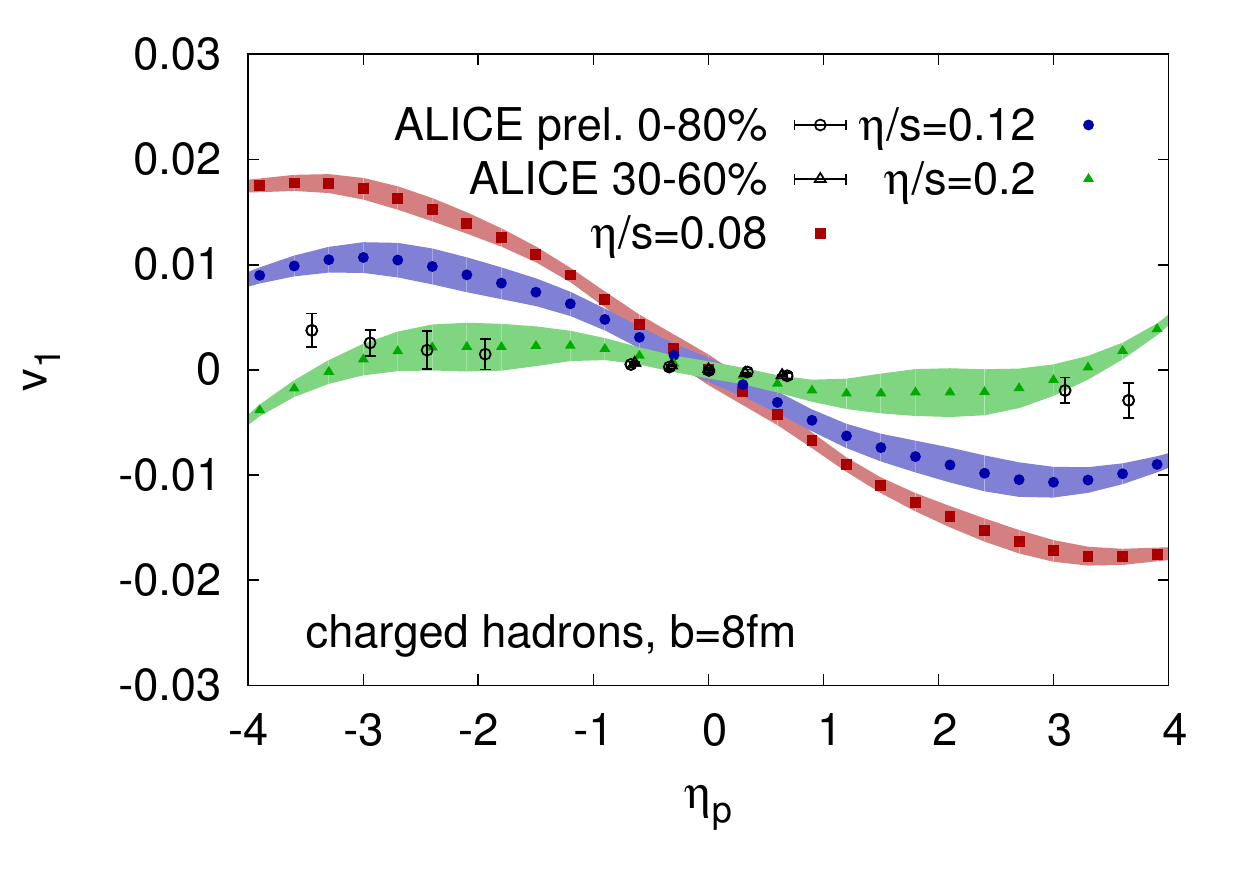}
\par\end{centering}

\protect\caption{Directed flow $v_{1}$ as a function of pseudo-rapidity
$\eta_p$ for 2.76 TeV collisions at LHC with $b=8\,{\rm fm}$
using $\eta/s= 0.08$, $0.12$, and $0.2$.
While a direct comparison with experimental data is difficult
without a proper event-by-event analysis, the shape around midrapidity
matches preliminary ALICE data for the 0-80\% centrality class \cite{Selyuzhenkov:2011zj,Gyulnara:2014cra} and published data for 30-60\% centrality \cite{Abelev:2013cva} quantitatively. 
The bands describe uncertainties in $v_1$ stemming from numerical inaccuracies in the freeze-out surface finding. \label{fig:v1-lhc}}
\end{figure}

One of the main motivations of this work was to study the rapidity
dependence of the directed flow, which we show in Fig.\,\ref{fig:v1} and Fig.\,\ref{fig:v1-lhc}
for collisions at RHIC (200 GeV) and LHC (2760 GeV), respectively. 
Here we used an impact parameter of 8 fm to simulate approximately 35\% central  collisions.

We have computed the event plane $\psi_1$ at forward rapidities and then evaluated 
\begin{equation}
  v_1 = \langle \cos(\phi-\psi_1)\rangle\,,
\end{equation}
where $\phi$ is the azimuthal angle of charged hadrons in momentum space and the average
$\langle \cdot\cdot\cdot \rangle$ is over the particle momentum distributions.

Interestingly, this quantity turns out to
be very sensitive to the $\eta/s$ ratio, and can as such possibly
be a good probe of the viscosity. The shape of the curve matches 
the STAR result quantitatively close to midrapidity, which can be seen as a partial
success of the combination of the holographic rapidity profile and
the LCOM description presented above. At larger rapidities, one should not trust the results for $v_1$,
because the multiplicity distribution is not described well in this regime.
The bands shown in Figs.\,\ref{fig:v1} and \ref{fig:v1-lhc} describe uncertainties from inaccuracies in the
determination of the freeze-out surface. They are obtained by mirroring the result around the origin,
which is a symmetry of the hydrodynamic initial condition.

\begin{figure}
\begin{centering}
\includegraphics[width=9cm]{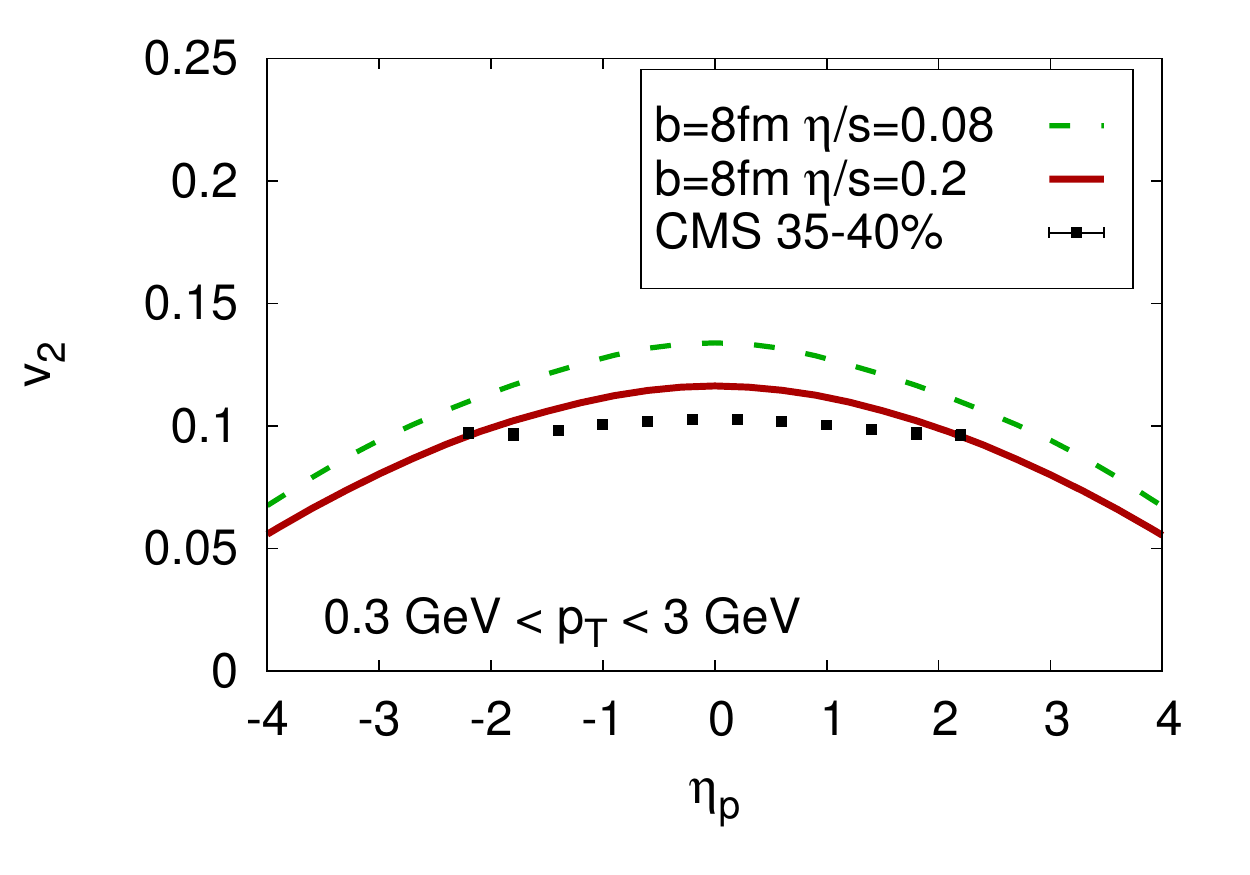}
\par\end{centering}

\protect\caption{Elliptic flow $v_{2}$ as a function of pseudo-rapidity
$\eta_p$ for 35-40\% central collisions at $2.76\,{\rm TeV}$ compared to experimental data
from the CMS collaboration \cite{Chatrchyan:2012ta}.
We compare two different values of $\eta/s$, 0.08 and 0.2.\label{fig:v2}}
\end{figure}

Fig. \ref{fig:v2} shows the elliptic flow as a function of pseudo-rapidity,
compared to experimental data from the CMS collaboration \cite{Chatrchyan:2012ta}. 
Similar to the result for multiplicity vs. pseudo-rapidity (Fig. \ref{fig:dndeta}),
the pseudo-rapidity dependence of $v_2$ is stronger than in the experimental data,
while the overall magnitude is close to the experimental data when using $b=8\,{\rm fm}$ and $\eta/s=0.2$.

We wish to stress that the presented model is very constrained compared to competing
models, and basically has no free parameters apart from the energy
density rescaling. The only input for the initial stage is the radius
and energy of the incoming nuclei and the equation of state of QCD
($e/T^{4}\approx 12$). Note, however, that adding a more refined transverse
energy profile with fluctuations in the future will introduce extra scales relating
to those fluctuations.

Nevertheless, when directly comparing with data our simple model has
two clear problems. Firstly, to get a reasonably $dN/d\eta_p$ at mid-rapidity we artificially reduced the
initial energy density by a factor 20 (6) for LHC (RHIC) collisions,
which are large factors.
Secondly, the $dN/d\eta_p$ spectrum
is still too narrow, even though the width is much more in line with
experimental data than the initial strong coupling profile found in
\cite{Casalderrey-Solana:2013aba}. These problems can be partly
attributed to the fact that QCD has an intermediate coupling strength, which
will intuitively lead to more particles at higher rapidity than in the strong coupling limit presented here.
Also, it may be possible that holography is better thought of as providing a description
of the (soft) gluons, which carry only part of the energy of the nucleus. The valence quarks 
perhaps require a different picture
\footnote{This is related to the holographic finding that all energy and also
all baryon charge initially ends up in the plasma at relatively
small rapidity \cite{Casalderrey-Solana:2014charge}.
}. Furthermore, importantly, including fluctuations
also partly resolves both problems. The reason is simple: fluctuations
will cause QGP to end up at high positive or negative rapidities, thereby widening the rapidity profile.
This widened rapidity profile will also give a lower total multiplicity, since the total input energy is fixed and particles
at higher rapidity carry much more energy.

Lastly, this work opens up many possibilities for further research.
First of all the holographic model can be improved to model heavy
ion collisions more realistically. Including finite coupling
corrections will be an important step in this direction (see i.e. \cite{Steineder:2012si}).
It will also be essential to include event-by-event fluctuations.
Interestingly, these studies will rely on a particularly rich set of physics,
from numerical general relativity to relativistic hydrodynamics to
particle decays, which will allow to make a quantitative comparison
of results obtained using holography to part of the experimental data
at RHIC and LHC.

\noindent \textbf{Acknowledgments. } We thank Jorge Casalderrey-Solana, Paul Chesler, Michal
Heller, Krishna Rajagopal and Paul Romatschke for useful discussions. WS is supported by the U.S. Department of
Energy under DOE Contract No. DE-SC0011090. BPS is supported by the U.S. Department of
Energy under DOE Contract No. DE-SC0012704. BPS gratefully acknowledges a DOE Office of Science Early Career Award.

\bibliographystyle{apsrev}
\bibliography{draft2.7}

\end{document}